\documentclass[a4paper]{article}
\usepackage{amsmath}
\usepackage{amssymb}
\usepackage{amstext}
\usepackage{amsthm}
\usepackage{wrapfig}
\usepackage{url}
\usepackage{authblk}
\usepackage[dvipsnames]{color}
\usepackage{graphicx}

\DeclareMathOperator*{\argmin}{arg\,min}

\begin{document}
\newcommand{\measurement}{\mathbf{y}}
\newcommand{\signal}{\mathbf{x}}
\newcommand{\reps}{\mathbf{c}} 
\newcommand{\repsHat}{\hat{\reps}} 

\title{Deep Compressed Learning for 3D Seismic Inversion}

\author[1]{Maayan Gelboim} 
\author[1]{Amir Adler} 
\affil[1]{Braude College of Engineering, Karmiel, Israel.}
\author[2]{Yen Sun}
\author[2]{Mauricio Araya-Polo} 
\affil[2]{TotalEnergies EP Research \& Technology US., Houston, Texas.}

\maketitle

\section{Summary}
\vskip -5pt
We consider the problem of 3D seismic inversion from pre-stack data using a very small number of seismic sources. The proposed solution is based on a combination of compressed-sensing and machine learning frameworks, known as \textit{compressed-learning} \cite{calderbank2009compressed,calderbank_jafarpour_2012,zisselman2018compressed,mdrafi2021compressed,9439820,9841016}. The solution jointly optimizes a dimensionality reduction operator and a 3D inversion encoder-decoder implemented by a deep convolutional neural network (DCNN). Dimensionality reduction is achieved by learning a sparse binary sensing layer that selects a small subset of the available sources, then the selected data is fed to a DCNN to complete the regression task. The end-to-end learning process provides a reduction by an \textit{order-of-magnitude} in the number of seismic records used during training, while preserving the 3D reconstruction quality comparable to that obtained by using the entire dataset.
\section{Introduction}
\vskip -5pt
Earth models can be used for many purposes, such as: seismology studies, geohazard estimation, hydrocarbon exploration and CO$_2$ sequestration. When used for CO$_2$ sequestration, subsurface models are critical inputs to analysis and decisions. Reconstruction of earth models is highly complex, involving a very large number of variables and large-scale datasets. 
Reconstructing 3D geological models by seismic inversion is a highly challenging task due to the huge amounts of seismic records, and the very-high computational cost of iterative numerical solution of the wave equation, as required by traditional algorithms such as Full Waveform Inversion (FWI). 
Just in terms of data, thousands seismic experiments are required for 3D earth model reconstruction of a typical exploration block, leading to Terabytes of recorded data. 

Since the direct computation (and lack of key parameters) of the most sophisticated approximation of the wave equation is not tractable, it is common to use the inverse approach. Seismic velocity (density or others key parameters) inversion computes a complete 3D velocity model ($\hat{\textbf{m}}$) of a certain target area, from recorded seismic data $\mathbf{d_r}$, \textcolor{black}{and it can be summarized as}:
\begin{equation}
\hat{\textbf{m}}=\textbf{G}^{-1}(\mathbf{d_r}),    
\end{equation}
 where $\textbf{G}^{-1}$ is the inversion operator. Seismic inversion problems \cite{Schuster17} are ill-posed, e.g., the solution is non-unique and unstable in the sense that small noise variations may alter the solution significantly. The solution of 3D seismic inversion problems using deep learning (DL) is an emerging field of research, see for example  \cite{gelboim2022encoder,tan2021deep,zeng2021inversionnet3d}, which were motivated by state-of-the-art results obtained by DL for the 2D case \cite{9363496,Yang19,araya2020fast,park2020automatic,zhang2022regularized,9808328,fabien2020seismic,zhu2022integrating}. In this study we address the problem of 3D velocity inversion, while performing shots selection \cite{araya2015greedy} to reduce significantly the computational load of the inversion process as well as required storage for acquired data. The proposed solution is demonstrated in an area with surface dimensions of 4.5km $\times$ 1km, which is modeled with $58$ seismic experiments for 3D model reconstruction. Nevertheless, our solution is scalable to larger area dimensions and higher numbers of experiments. The target applications are: improve repeatability of seismic surveys, reduction of necessary sources for inversion algorithms (ML or other), acquisition design improvement, and inversion results explainability. Subsequent surveys of a given 4D setup can concentrate on selected sources, therefore reducing noise/mismatches coming from not selected sources. When new data is acquired the corresponding training can focus on the selected sources only, thus reducing significantly the memory and computing costs. Eventually, the collections of selected shots can be also use with other more traditional algorithms such as FWI. Knowing the most relevant sources can be incorporated in future acquisition designs, thus reducing costs. Finally, convergence and results analysis can improve by knowing exactly what sets of sources contribute the most.
\section{Theory and Methods}
\vskip -5pt
\begin{table}
\small
\caption{Proposed Joint Sensing and Inversion Architecture}
\centering
\begin{tabular}{|l|l|l|l|}
\hline

\textbf{Block} & \textbf{Layer} & \textbf{Unit} &  \textbf{Comments}  \\
\hline
Input &  0  & Seismic Cube  & $58 \text{ Cubes}\times (192 \times 64 \times 496$)          
\\
\hline
Sensing & 1 &  Binarized Sensing Layer  & One bit per shot  (i.e. cube)\\ \hline
 Enc1 &  2  & Conv3D(32, ($5\times 5\times5$),ReLU)   &  + InstanceNormalization \\
      &  3  & Conv3D(32, ($5\times 5\times5$),ReLU)   &  + InstanceNormalization \\
      &  4  & MaxPool3D   & + Dropout(0.2)      \\\hline
Enc2 & 5-7 & Enc1(64) & \\\hline
Enc3 & 8-10 & Enc1(128) &  \\\hline
Enc4 & 11-13 & Enc1(256) & \\\hline
Enc5 & 14-15 & Enc1(512) & without MaxPool3D\\\hline
Dec1 &  16  & ConvTrans3D(256, ($2\times 2\times2$))      &  + InstanceNormalization       \\

           &  17  &  Conv3D(256, ($5\times 5\times5$))     &  + InstanceNormalization\\

           &  18  & Conv3D(256, ($5\times 5\times5$))     &   + InstanceNormalization\\
           & & & ReLU in all Conv. layers\\\hline
Dec2 & 19-21 & Dec1(128) & \\\hline
Dec3 & 22-24 & Dec1(64) & \\\hline
Dec4 & 25-27 & Dec1(32) & \\\hline
 & 28 &  Conv3D(1, ($1\times 1\times1$),ReLU) & Final reconstruction layer\\\hline
Output &  29  & Velocity Model  & $192 \times 64 \times 496$         grid points\\
\hline
\end{tabular}
\label{tab:CNN-ENC-DEC}
\end{table} 
Compressed Learning (CL)  facilitates solutions of machine learning tasks, such as classification and regression, from compressed sensing (CS) measurements, therefore bypassing the standard requirement of full sample rate signal acquisition (either temporal or spatial). In the following we briefly explain CS and CL concepts, followed by a detailed description of the proposed solution.\\

\noindent\textbf{Compressed Sensing:}
Given a signal $\signal \in \mathbf{R}^N$, an $M \times N$ sensing matrix $\Phi$ (such that $M \ll N$) and a measurements vector $\measurement = \Phi \signal$, the goal of CS is to recover the signal from its measurements. The sensing rate is defined by $R=M/N$, and since $R \ll 1$ the recovery of $\signal$ is not possible in the general case. According to CS theory \cite{1614066}, signals that have a sparse representation in the domain of some linear transform can be exactly recovered with high probability from their measurements: let $\signal = \Psi \reps $, where $\Psi$ is the inverse transform, and $\reps$ is a sparse coefficients vector with only $S \ll N$ non-zeros entries, then the recovered signal is synthesized by $ \hat{\signal} = \Psi \repsHat$, and $\repsHat$ is obtained by solving the following convex optimization program:
\begin{equation}
 \repsHat  = \argmin_{\reps{'}} \left\|\reps{'}\right\|_1 \text{ subject to } \measurement = \Phi \Psi \reps{'},
\end{equation}
\noindent where $\left\|\alpha\right\|_1$ is the $l_1$-norm, which is a convex relaxation of the $l_0$ pseudo-norm that counts the number of non-zero entries of $\alpha$. The exact recovery of $\signal$ is guaranteed with high probability if $\reps$ is sufficiently sparse and if certain conditions are met by the sensing matrix and the transform. While CS was originally developed for general sensing matrices, it was later extended \cite{8052513} to the special case of binary sensing matrices.\\

\noindent\textbf{Compressed Learning:}
CL was introduced by \cite{calderbank2009compressed}, which theoretically proved that direct inference from the compressive measurements $\measurement = \Phi \signal$ is possible  with high classification accuracy. In particular, this work provided analytical results for training a linear Support Vector Machine (SVM) classifier in the CS domain. CL implementations based on deep learning were introduced by \cite{adlercompressed,zisselman2018compressed}. 

In this work we implemented a compressed learning method, which \textit{jointly} optimizes a sparse binary sensing tensor $\Phi$ and a non-linear 3D inversion operator $\mathcal{R}(\Phi \odot \mathbf{d};W)$, parameterized by a coefficients matrix W, and $\odot$ is the element-wise multiplication operator. The shot-gathers data tensor $\textbf{d}$ is element-wise multiplied with the sensing tensor $\Phi$, which nulls most of the shots and effectively reduces the number of shot-gathers that are utilized as input for the inversion operator. The inversion operator is an encoder-decoder that maps a set of shot-gathers (sorted seismic records) to a 3D velocity model, implemented using a deep convolutional neural network (DCNN). 
The proposed method provides a solution to the following joint optimization problem:
\vskip -10pt 
\begin{equation}
\{\hat{\Phi},\hat{W}\}= \arg \min_{\phi,W}{\frac{1}{N}\sum_{i=1}^{N}\mathcal{L}(\mathbf{m_{i}},\mathcal{R}(\Phi \odot \mathbf{d_{i}};W),R,R_\phi)},
\end{equation}
\noindent where $\{\mathbf{d_{i}},\mathbf{m_{i}}\}_{i=1}^N$
is the training-set of $N$ pairs of shot-gathers $\mathbf{d_{i}}$ and corresponding 3D velocity-models (i.e. labels)  $\mathbf{m_{i}}$. The loss function $\mathcal{L}(\mathbf{m_{i}},\hat{\mathbf{m_{i}}},R,R_\phi)$ measures two misfit terms: (i) the mean absolute error (MAE) between the ground-truth velocity model (i.e. label) and the estimated one $\hat{\mathbf{m_{i}}}$, provided by the DCNN encoder-decoder reconstruction operator $\mathcal{R}(\bullet)$, whose input is the compressively-sensed subset of shot gathers, denoted by $\Phi \odot \mathbf{d_i}$; and (ii) the squared error between the target sensing rate $R$ $( 0<R<1)$ and the learned sensing rate $R_\phi$. The loss function is defined as follows:
\begin{equation}
 \mathcal{L}(\mathbf{m_{i}},\hat{\mathbf{m_{i}}},R,R_\phi) = \ \text{MAE}(\mathbf{m_{i}},\hat{\mathbf{m_{i}}}) + \lambda(R-R_\phi)^2,
\end{equation}
where $\lambda>0$ is a weight that  controls the trade-off between the two misfit terms, and 
\vskip -15pt
\begin{equation}
    \text{MAE}(\mathbf{m},\hat{\mathbf{m}})=\frac{1}{K}\sum_{j=1}^{K}|m_j-\hat{m}_j|,
\end{equation}
 \vskip -5pt
where $m_j,\hat{m}_j$ are the grid point entries of the (column-stacked) ground truth 3D model $\mathbf{m}$ and the inverted 3D model $\hat{\mathbf{m}}$, respectively (each model with $K$ grid points). 
\indent In this work we implemented and trained a 3D convolutional encoder-decoder, inspired by the 2D U-Net architecture \cite{UNET}, to learn the mapping from seismic data space to 3D models space (i.e. inversion). 
 The complete network architecture follows the implementation introduced by \cite{gelboim2022encoder} and detailed in Table \ref{tab:CNN-ENC-DEC}, with a total of 99M parameters. Note that during training the sensing layer and the subsequent layers, represented  by $\mathcal{R}(\bullet)$, are treated as a single deep network. However, once training is complete, the sensing layer is detached from the subsequent inference layers, and used for performing signal sensing. The input of the inversion operator $\mathcal{R}(\bullet)$, is therefore the second layer of the end-to-end learned network.
 We next discuss the construction of the sensing tensor  $\Phi$, which follows closely the concept of \textit{binarized neural networks}, presented in \cite{hubara2016binarized}. Let \begin{math} \phi_w\in R^{N_s\times 1}\end{math} be the real-valued (non-binary) weights of the trainable sensing layer, where $N_s$ is the number of shots. In order to convert $\phi_w$ to a binary vector, we employ a \textit{binarization} function, that hard-thresholds the real values using a fixed threshold, resulting in the binarized vector $\phi_b \in \{0,1\}^{N_s \times 1}$. Note that each bit in $\phi_b$ corresponds to one shot, where a value of `0' corresponds to discarding the shot and `1' corresponds to keeping the shot. The learned sensing rate is defined as follows:
 \vskip -10pt
 \begin{equation}
     R_\phi=\frac{1}{N_s}\sum_{i=1}^{N_s}{\phi_b(i)},
 \end{equation}
where $\phi_b(i)$ is the \textit{i}-th entry of $\phi_b$. The sensing tensor $\Phi$ performs gating of the input shot-gathers, according to the binarized values in $\phi_b$, by element-wise multiplication with the data tensor  $\mathbf{d}\in R^{N_s \times N_{rx} \times N_{ry} \times T}$, where $N_{rx}, N_{ry}, T$ are the number of horizontal-axis receivers, number of vertical-axis receivers and number of time-samples, respectively. Therefore, the sensing tensor $\Phi$ is required to have identical dimensions to $\mathbf{d}$ and it is constructed by replicating the binarized value per-shot in $\phi_b$ for all the receivers and time-sample entries, corresponding to that specific shot in $\Phi$. Since the derivative of the \textit{binarization} function has a discontinuity with an indefinite  value, it cannot be utilized for back-propagation. Therefore, the forward-pass employs the binarization function, whereas the backward-pass  utilizes the \textit{hard-sigmoid} function $\sigma_H(x)$, 
 defined as follows:
\begin{equation}
\sigma_H(x) =  \max(0, \min(1,\frac{x + 1}{2})),
\end{equation}
which is linear with slope 0.5 for $ -1\le x\le 1$  and has a well-defined derivative for all values of $x$, where $x=\phi_w(i)$ with $i=1,\dots ,N_s$.
\begin{figure}[ht] 
   \centering
  \includegraphics[trim={0.0cm 0 0.1cm 0},clip,scale=0.55]{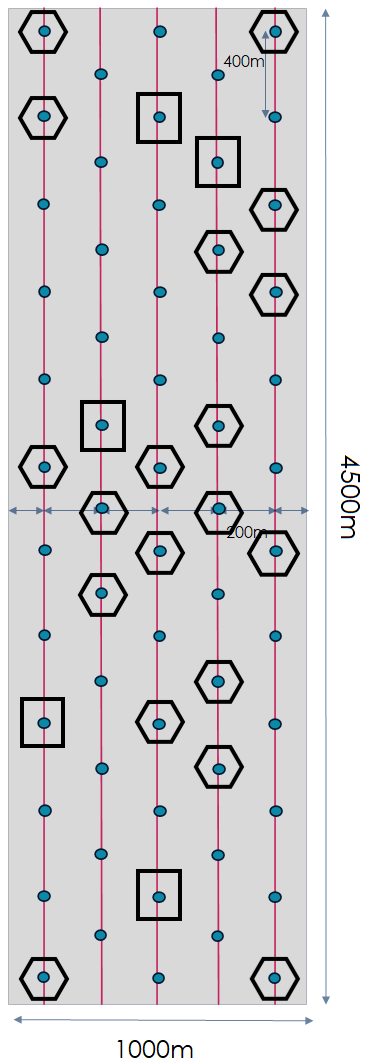}
    \caption{Geometry of the 58 shot positions (blue circles). The sensing layer selects only a fraction of the available 58 shots. The blue circles within hexagons represent the shots selected with $30\%$ of the original shots, and the ones within squares are selected with only $10\%$ of the original shots.}
    \label{fig:geometry}
\end{figure}
\section{Results and Observations}
\vskip -5pt
Performance of the proposed approach was evaluated by generating
2000 3D models of dimensions 4.5Km $\times$ 1.0Km $\times$ 4Km (length $\times$ width $\times$ depth), with variable thickness
velocity layers, including random salt geometries in $40\%$ of the models. All models were generated using the Gempy (\url{https://www.gempy.org/}) tool, for 3D geological modeling. Seismic surveys were simulated using an isotropic acoustic wave approximation (see \cite{meng2020minimod}), $58$ shots per model were arranged along 5 parallel columns, as illustrated in Figure \ref{fig:geometry}, with adjacent columns separated by 200m and evenly-spaced shots per column. Acquisition geometry was set to a uniform rectangular grid of receivers, separated by 15m in the horizontal and
vertical axes. All traces were further multiplied by a monotonically-increasing time-boosting function that
compensates the attenuation of wave reflections from the lowest geological layers by amplifying late-arrival time samples. We split the 2000 samples
(velocity models as labels and their corresponding traces as data) into 1600 training and 400 testing samples. We
initially trained the DCNN without compressed sensing (CS), using our baseline system that has no sensing layer and processes the entire collection of shots. To overcome the high computational requirements of this, we utilized the dimensionality reduction scheme, proposed in  \cite{gelboim2022encoder}, in which all shot-gathers are spatially averaged along the shots dimension, resulting in a single data cube denoted by $\bar{\mathbf{d}}(R_x,R_y,t)$, where $R_x,R_y$ are the indices of the receiver positions, and $t$ is time-sample index. The dimensionality-reduced data cube is defined as follows:
\begin{equation}
\label{eq:data-cube}
 \bar{\mathbf{d}}(R_x,R_y,t)=\frac{b(t)}{N_s}\sum_{S=1}^{N_s}{\mathbf{d}(S,R_x,R_y,t)},
\end{equation}
where $b(t)$ is a monotonically-increasing time-boosting function. Therefore, $\bar{\mathbf{d}}$ forms a single 3D input channel, thus significantly mitigating  the memory and computational requirements for training and inference of the proposed DCNN, as discussed in \cite{gelboim2022encoder}. We further trained the joint 
sensing and inversion  architecture, as detailed in Table 1, with target sensing rates of $30\%$ and $10\%$. Note that the spatial averaging along the shots dimension was not employed to the CS shots, and each shot was processed by a separate 3D convolutional layer.  Reconstructed velocity models' quality was
evaluated on the testing set, using the 3D Structural Similarity Index Measure (SSIM) \cite{1284395}, computed by averaging
the 2D SSIM's along the XY, XZ and YZ planes. 3D SSIM results 
evaluated on the held-out testing set were $0.9265$ out of $1.0$ (without CS), $0.9262$
(with $30\%$ CS rate) and $0.9225$ (with $10\%$ CS rate). Visual examples of reconstructed 2D slices are provided in figure 2(a), and of 3D models are provided in figure
2(b)-(c), clearly demonstrating the good velocity model building quality, even at a 10\% compressed sensing rate.
During training using $10\%$ of the sources is drastically more computationally efficient than using all of them, but it is more expensive than using the approach in Equation~\ref{eq:data-cube}. The advantage of using CS is that when a small set of sources is selected, the subsequent training and acquisition design will be more efficient. Further, pre-processing steps costs are reduced if one decide to use Equation~\ref{eq:data-cube} along with the CS selected shots, this combination yields the best solution in terms computational cost and reconstruction accuracy, plus helps to better understand the learning process in relation to the data used. 
\begin{figure*}
\hskip -0.9in
\includegraphics[width=1.5\columnwidth]{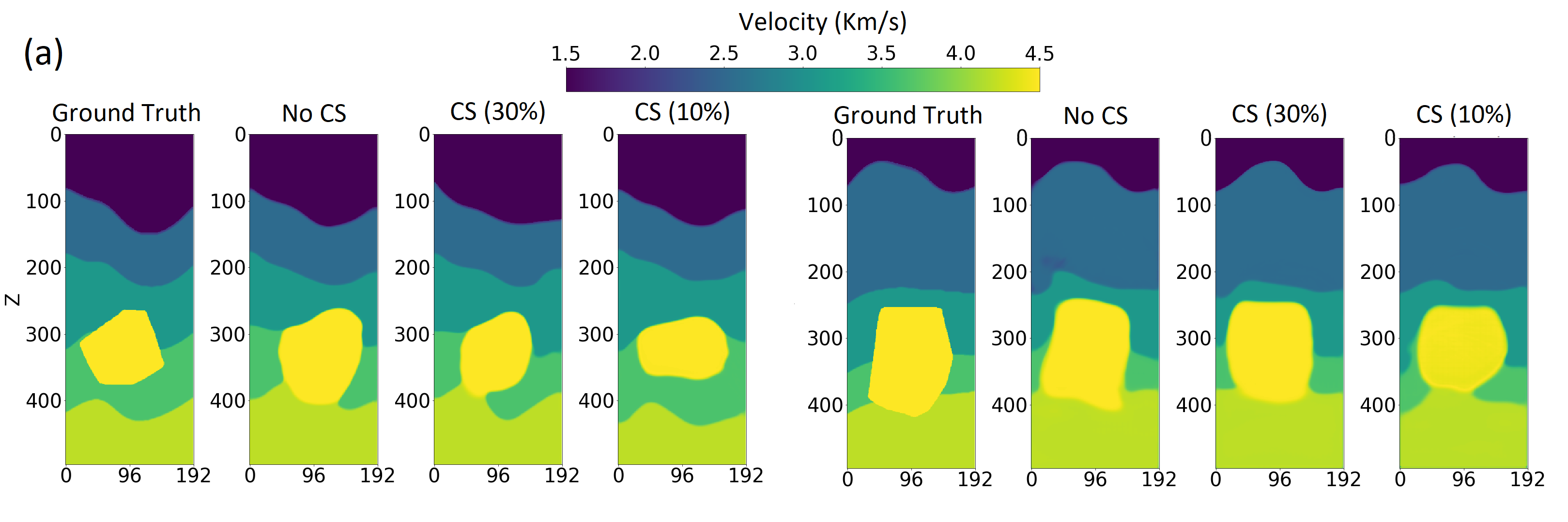}\\
\vskip 0.5cm
\hskip -0.3in
\includegraphics[scale=0.23]{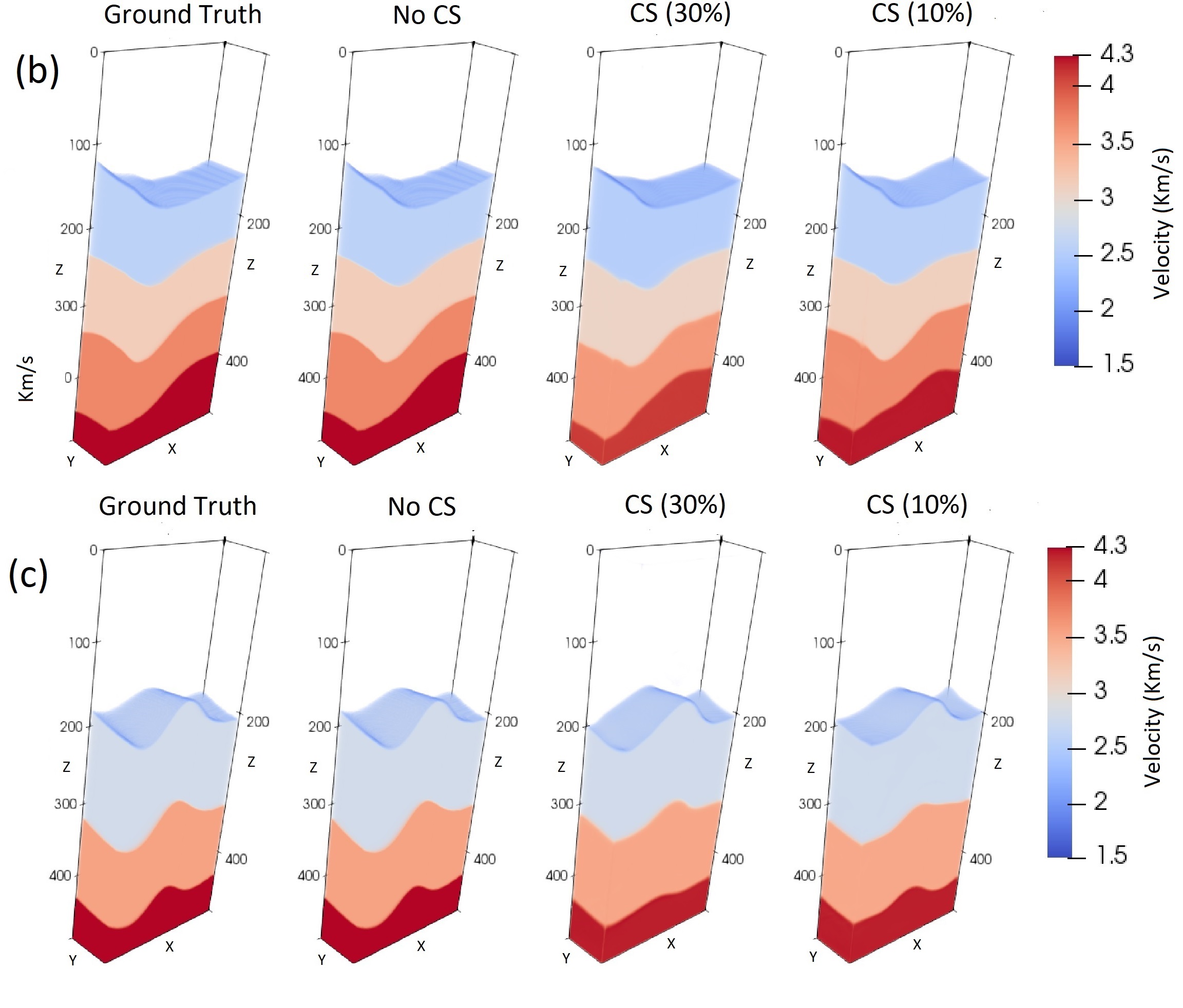}
\caption{ (a) Samples of 2D slices from 3D models testing-set, illustrating high-quality reconstructions of
the variable thickness layers interfaces, and salt geometries, at sensing rates of $30\%$ and $10\%$. (b) and (c) Samples of reconstructed 3D models illustrating high-quality reconstructions at sensing rates of $30\%$ and $10\%$.}
\label{fig:CS_results}
\end{figure*}

\section{Conclusions}
Seismic inversion for reconstructing 3D geological structures is a highly challenging task due to  the huge amounts of acquired seismic data, and very-high computational load required for iterative numerical solutions of the wave equation, as required by Full Waveform Inversion. We presented a solution that jointly optimizes the compressed sensing operator with the 3D inversion network, leading to an order-of-magnitude reduction of the number of required shot-gathers for high-quality 3D inversion. The solution  facilitates lower memory and computational requirements, and can be further extended to simplify and optimize the acquisition setup geometry.

\section{Acknowledgments}
The authors acknowledge TotalEnergies EP Research and Technology US, for supporting this work and allowing its publication.

\bibliographystyle{ieeetr}
\bibliography{Main}
\end{document}